\documentclass[aps,notitlepage,reprint,footinbib,citeautoscript,onecolumn]{revtex4-1} 
\usepackage{graphicx} 
\usepackage{subfigure}
\usepackage{dcolumn} 
\usepackage{bm} 
\usepackage{float} 
\usepackage{bbm} 
\usepackage{enumitem,xcolor}
\usepackage{color}
\usepackage{amsfonts}
\usepackage{xcolor}
\usepackage{lmodern} 
\usepackage{natbib}
\usepackage{comment}
\usepackage{amsmath} 
\usepackage{amsmath,amssymb}
\usepackage{natbib}
\usepackage[margin=2cm]{geometry}
\usepackage{tikz}
\usepackage{ulem}
\usepackage{lineno}
\usetikzlibrary{matrix}

\newcommand\bi{\begin{itemize}}
\newcommand\ei{\end{itemize}}

\usepackage[colorlinks=true,citecolor=blue]{hyperref}
\hypersetup{colorlinks=true,citecolor=blue,linkcolor=blue,urlcolor=black}

    \usepackage{geometry}

    \usepackage{tikz} 
    \usetikzlibrary{shapes,arrows,calc}
    \usetikzlibrary{datavisualization}
    \usetikzlibrary{datavisualization.formats.functions}
    \usepackage{pgfplots} 
    \usepackage{tikz}
\usetikzlibrary{matrix}  

\usepackage[colorlinks=true,citecolor=blue]{hyperref}

\begin{document}
\title{Physics of liquid crystals in cell biology}

	\author{Amin Doostmohammadi}
	\affiliation{The Niels Bohr Institute, University of Copenhagen, Copenhagen, Denmark}
	\email{doostmohammadi@nbi.ku.dk}
	\author{Benoit Ladoux}
	\affiliation{Institut Jacques-Monod, CNRS and Université de Paris, France}
	\email{benoit.ladoux@ijm.fr}

\maketitle

The last decade has witnessed a rapid growth in understanding of the pivotal roles of mechanical stresses and physical forces in cell biology. As a result an integrated view of cell biology is evolving, where genetic and molecular features are scrutinized hand in hand with physical and mechanical characteristics of cells. Physics of liquid crystals has emerged as a burgeoning new frontier in cell biology over the past few years, fueled by an increasing identification of orientational order and topological defects in cell biology, spanning scales from subcellular filaments to individual cells and multicellular tissues. Here, we provide an account of most recent findings and developments together with future promises and challenges in this rapidly evolving interdisciplinary research direction.

~\\
\noindent
\fcolorbox{red}{white}{\begin{minipage}{\textwidth}
{
{\bf Glossary}
\begin{itemize}
    \item {\bf Active liquid crystal}, Liquid crystal driven constantly away from equilibrium by local injection of energy from its constituent elements.
    \item {\bf Mechanotransduction}, Conversion of mechanical cues received by cells into biochemical signals within the cells.
    \item {\bf Liquid crystal}, Intermediate phase between solid and liquid that flows like a liquid but maintains some of the ordered structure of crystals.
    \item {\bf Polar alignment}, Alignment of particles with a distinct head and tail orientations along the same direction.
    \item {\bf Nematic alignment}, Alignment of rod-like particles which present head-tail symmetry in their orientation. Nematic ordering can be obtained in two ways, either in systems where polar objects are parallel with random head-tail orientations or in systems where particles are themselves head-tail symmetric.
    \item {\bf Topological defects in orientational alignment}, Imperfections in liquid crystalline organisation.
    \item {\bf Topological transition}, Cell shape remodeling within tissues based on cellular forces that lead to fluid/solid, jammed/unjammed behaviors.
    \item {\bf Reynolds Number}, Dimensionless quantity used in hydrodynamics that represent the ratio between inertial forces and viscous forces. Laminar flows occur at low Reynolds number, whereas turbulent flows appear at high Reynolds number.
\end{itemize}
}
\end{minipage}}

\section*{The Rising (orientational) Order in Cell Biology}

From the earliest stages of development, where few stem cells self-organise into delicate well-structured aggregates to form various organs~\cite{Xavier2019}, all the way to the organ repair and wound healing, where cells actively coordinate their behavior to close gaps and scars in the tissue~\cite{Solon2009,Vedula2014, Park2017}, multiple genetic, chemical, and mechanical factors work together to control the spatio-temporal order in cells~\cite{Mao2016}. Importantly, the breakdown of the order is at the root of tissue malfunctioning, impaired healing, and disease initiation~\cite{TeBoekhorst2016,Wu2015}.
The order is not only in the spatial arrangement of cells, but also comprises meticulous orientational alignment to govern the direction of cell division~\cite{Wyatt2015,Heisenberg2013} and cell migration~\cite{Vedula:2012aa}, as well as directional transmission of mechanical forces and biochemical signals between the cells~\cite{Trepat2018}. For example, coordinated cellular migration and directed cell division events determine the elongation of embryos at the early stages of morphogenesis~\cite{Campinho2013}, deciding where particular parts of the body will be located. Similarly, force transmission through cell-cell contacts provides a directional guidance for the transfer of biochemical signals into specific locations within the tissue~\cite{Sebbagh2009,Das2015,Bays2017,Boocock2021}. These processes rely on the active nature of biological systems composed of living particles capable of continuously converting chemical energy into mechanical work~\cite{Prost2015}. Interaction of active particles can give rise to collective motion and the emergence of order. Remarkably, the orientational order is established and maintained across a range of scales, from multicellular organisation at the tissue level~\cite{Barlan2017,Notbohm2016,Jain2020}, to the alignment of individual cells~\cite{Zemel2010,Trichet2012}, and all the way down to subcellular filaments that form the skeleton of the cells as exemplified by both {\it in vitro} and {\it in vivo} experiments~\cite{Sanchez2012,Jalal2019,Chen2019}. These emergent properties led to an analogy with a well-known phase of physical systems called ``liquid crystals" (Box 1) and in particular, nematics~\cite{Saw2018Review}. Other types of liquid crystals including cholesterics are important to describe biological processes (reviewed elsewhere~\cite{Mitov2017}) but we mainly focus here on nematics. Passive liquid crystals thus display the general features of organization order, the local order breakdown through the formation of topological defects (Box 2), whereas other aspects are inherent to active liquid crystals including the emergence of active stresses, the formation of those defects and the movement of the defects. Here, we review how the emergence of orientational order and its breakdown are generic themes in fundamental processes in cell biology and provide an universal mechanism for choreographing the directional passage of mechanical and biochemical information in time, and across different scales.\\

\begin{figure}[h]
\centering
  \includegraphics[width=0.9\linewidth]{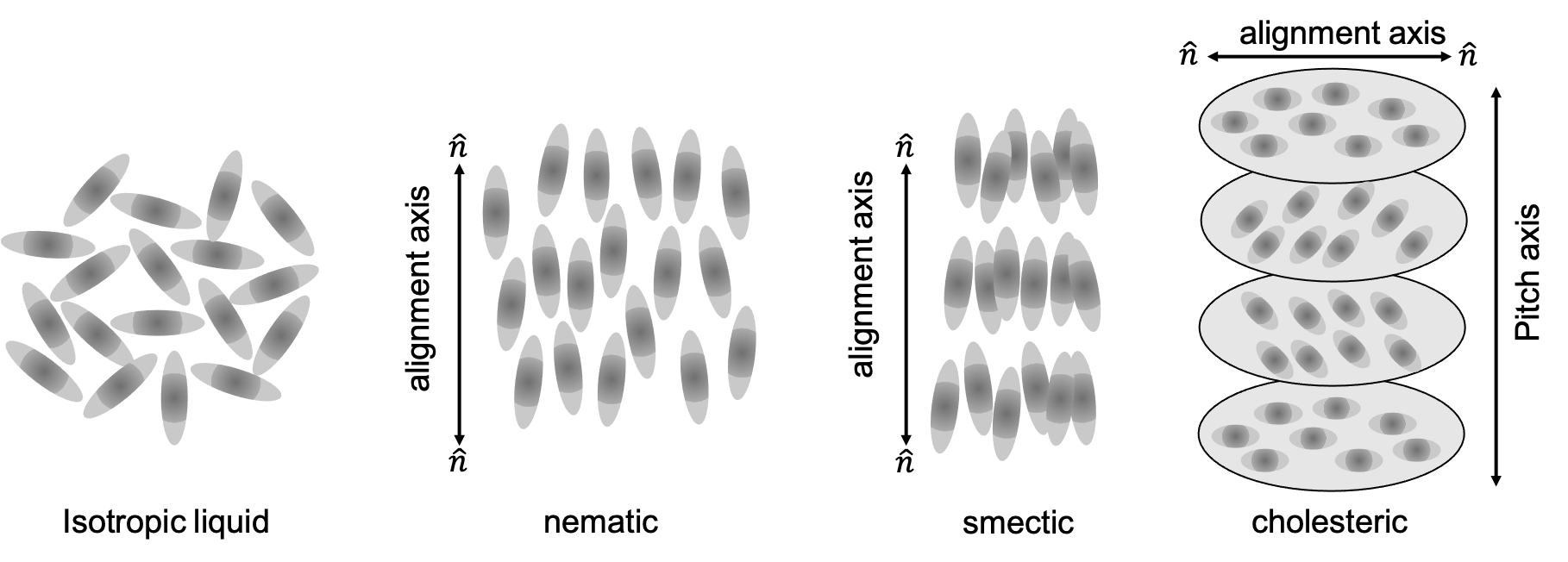}
  \caption{Different liquid crystalline phases.}
  \label{fig:lc}
\end{figure}

\noindent
\fcolorbox{red}{white}{\begin{minipage}{\textwidth}
{
{\bf BOX~1. Primer on Liquid Crystals}\\
Liquid crystals are an intermediate phase of matter between solids and liquids. In a solid constituent atoms organise in well-structured crystalline lattices giving them high degree of positional and orientational ordering. In a liquid both positional and orientational ordering are absent. As an intermediate phase of matter, the constituent elements of a liquid crystals organise in such a way that orientational order is maintained, but the positional order is lacking. This structural arrangement endows liquid crystals with peculiar properties: liquid crystals can flow like a liquid, but resist deformations like a solid.\\
The liquid crystal phase itself can exist in different sub-classes. In a {\it nematic} (from the Greek word for `thread-like') liquid crystal the molecules align (on average) along a common axis, but are free to drift around while keeping this orientational order (Fig.~\ref{fig:lc}). In a {\it smectic} (`soap-like') liquid crystal additional degree of order emerges: the molecules organise into layers along a particular direction such that they are liquid-like within each layer, but they all point (on average) parallel to the plan normal to the layer (Fig.~\ref{fig:lc}). A {\it chiral} (nematic or smectic) liquid crystal is formed when molecules posses helicity and periodically rotate along the alignment axis (Fig.~\ref{fig:lc}).
}
 \end{minipage}}

\section*{Physics of Liquid Crystals in Subcellular Organisation}

Within an individual cell, complex networks of filamentous protein structures shape the cell cytoskeleton. The cytoskeleton extends from the nucleus to the cell membrane and is integral in maintaining cell shape and modulating its response to external stimuli~\cite{Fletcher2010}. The main building blocks of the cell cytoskeleton are actin filaments, microtubule (made of protein tubulin), and intermediate filaments. As groups of these elongated filaments assemble together within a cell, various forms of orientational order emerge, sharing ordering properties of liquid crystals (see {\bf Glossary}). For example, {\it in vitro} reconstitution of microtubule~\cite{Decamp2015} and actin filament networks~\cite{Schaller10,Zhang2018} both demonstrate the emergence of nematic orientational order. Moreover, three-dimensional organisation of actin filaments-myosin motor protein mixtures in filopodia show chiral ordering that contributes to the helical rotation of the filopodia~\cite{leijnse2020filopodia}. Recently, the emergence of smectic order has also been observed in stacks of myosin-II filaments in epithelial cells~\cite{hu2017long}. Finally, the mechanical interplay between cell shape and cytoskeleton organization relies on nematic ordering of actin filaments at the single cell level, particularly important for rigidity sensing mechanisms ~\cite{Zemel2010,Gupta2015,Schakenraad2020,Doss2020}. The physical laws governing phase transitions and dynamics of liquid crystals have thus been reported in various subcellular processes, not only {\it in vitro} but also in living cells, as described below.

\subsection*{\it{Isotropic-nematic transition in actin filaments as a hallmark of cell polarization}}
It is shown through {\it in vitro} experiments that culturing cells on substrates with incrementally increased stiffness results in a sharp transition from circular, non-motile phenotype on soft substrates to elongated cells with prominent actin stress fibers on stiffer substrates~\cite{Prager-Khoutorsky:2011,Gupta2015}. A similar transition has been observed on substrates with modulated viscoelastic properties~\cite{Missirlis2020}. The sharp transition coincides with the isotropic-to-nematic transition of the actin cytoskeleton at a critical stiffness, below which the actin organisation is isotropic leading to circular shape of the cell, while above the transition point actin stress fibers self-organise into nematically ordered structures resulting in polarised elongated cells~\cite{Doss2020}. Importantly, the isotropic-to-nematic transition is shown to crucially depend on the activity of the cell cytoskeleton and its ability to generate contractile stresses. This distinguishes the transition from isotropic-to-nematic transition in passive liquid crystals, where thermodynamic ordering mechanisms such as increased density or reduced temperature govern the transition to a nematically ordered state. As such, the cell cytoskeleton shows properties of active liquid crystals~\cite{Gupta2015,Schakenraad2020}, where continuous injection of energy by the constituent elements - here through motor proteins - constantly drives the filaments away from thermodynamic equilibrium and the activity-induced ordering governs the transition from isotropic to nematic phase~\cite{Prost2015}.

\subsection*{\it{Polar versus Nematic Order in Subcellular Filaments}}
The presence of nematic order demarcates an emergent orientational order that is predominated by the alignment of elongated filaments along a particular axis without any preferable head-to-tail direction (polarity). In other words, nematic order along the direction $\hat{\bf n}$ is head-tail symmetric and thus does not change by the transformation $\hat{\bf n}\rightarrow-\hat{\bf n}$ (Fig.~\ref{fig:pol-nem}), meaning that the orientation order is equivalent along the head or tail direction. This implies that even polar filaments such as actin or microtubule that possess positive and negative ends can in principle self-organise into nematic structures if on average they align together without pointing into a specific direction, as has been reported in various cytoskeletal constructs including actin stress fiber organization~\cite{Vignaud2021,Zhang2018,Lehtimaki2021,Needleman2017}. However, such polar entities can in principal also self-organise into structures with polar orientational order, in which not only they align together, but also they predominantly point towards their positive ends as shown, for instance, in protrusive cellular structures such as filopodia and lamellipodia~\cite{Blanchoin2014}. This immediately raises a question of when and how one form of order, namely polar or nematic, dominates, whether it is even possible to form structures with mixed symmetry~\cite{amiri2021half} and eventually, how one structure can dynamically evolve into the other one to drive cell migration processes ~\cite{Yamashiro2014}.

Recent works have indeed taken first steps towards addressing these questions. It is shown that tuning the strength of nematic interactions between actin filaments by adding small amounts of PEG molecules can result in the emergence of polar waves co-existing with nematic bands in actomyosin motility assays~\cite{huber2018emergence}. Interestingly, experiments on microtubule-kinesin motor systems have also demonstrated the possibility of co-existing phases of polar and nematic alignment of microtubules with a governing role in forming the bipolar mitotic spindle in cells~\cite{roostalu2018determinants}. Importantly, such polar and nematic ordering of microtubules are shown to be controlled by two active features: microtubule growth speed to motor speed ratio, and the motor number to microtubule number ratio, which again highlights the importance of active, non-equilibrium factors in determining cytoskeletal structure. As such, a new perspective on subcellular network organisation is surfacing, in which the orientational order is an emergent property of the cell cytoskeleton that controls the network structure by switching between various forms of orientational order~\cite{huber2018emergence,roostalu2018determinants,amiri2021half}. 

\subsection*{\it{Topological Defects in systems of subcellular filaments}}
\begin{figure}[h]
\centering
  \includegraphics[width=0.9\linewidth]{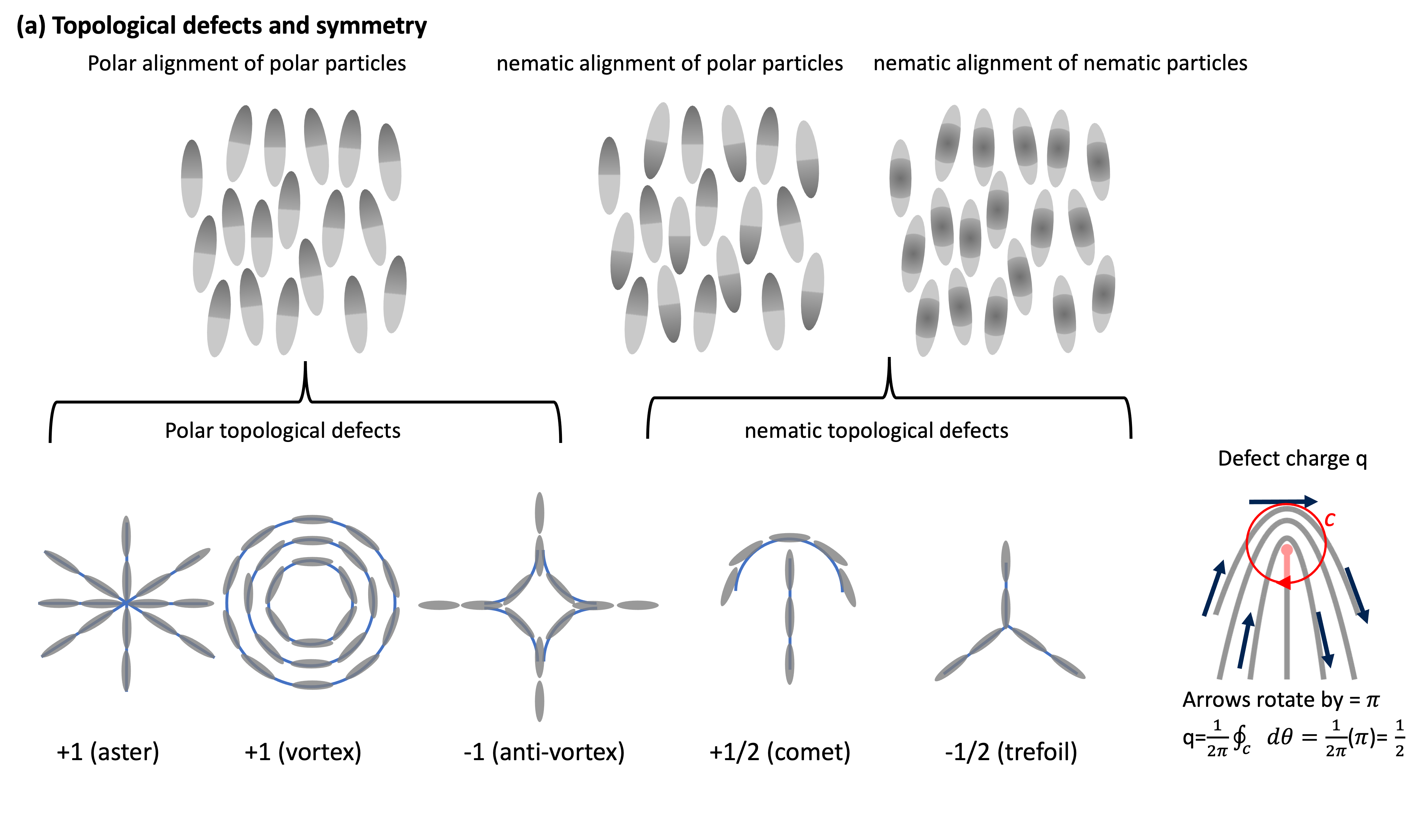}\\
  ~\\
  \includegraphics[width=0.9\linewidth]{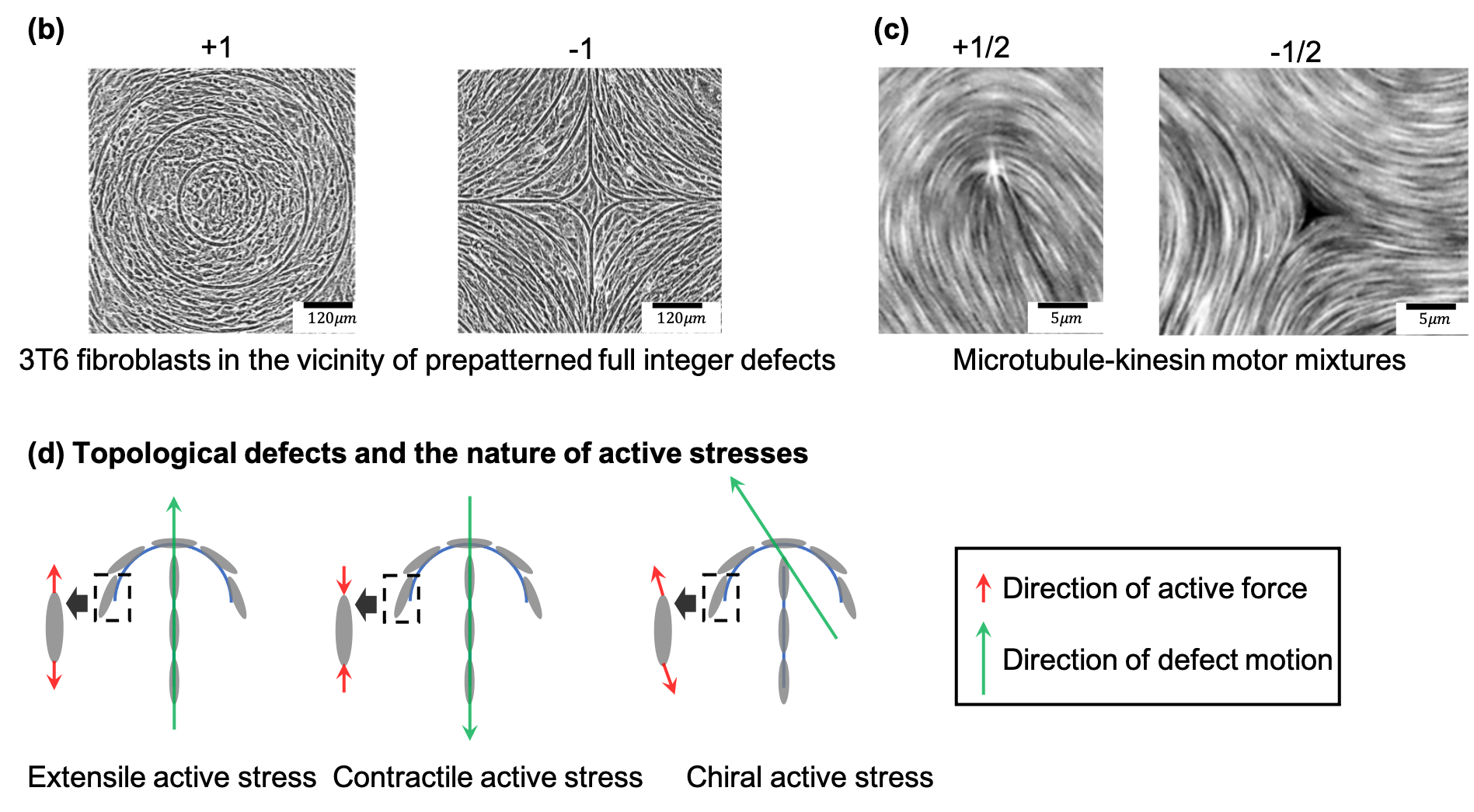}
  \caption{The form of topological defects is related to (a-c) the symmetry of the alignment of constituent particles of the system, and (d) the nature of active stresses that particles generate. In (a) an example of the calculation of defect charge around a $+1/2$ defect is shown: over a full, $2\pi$ clockwise rotation around the defect core (marked by the closed path {\it c}) the arrows that follow the orientation field rotate by $\pi$ in the clockwise direction and as such the defect has the charge of $+1/2$. Figures in (b) and (c) are adapted from~\cite{endresen2021topological,Doost18}, respectively.}
  \label{fig:pol-nem}
\end{figure}

An almost inevitable consequence of the establishment of orientational order is the possibility of order breakdown. Indeed, the domains of parallel orientation in any nematic material cannot indefinitely extend in space and are characterized by finite orientational correlation lengths. Bends and splays can thus emerge in the material that distort the orientations, and high bend and splay can lead to the local misalignment of the anisotropic particles. Such singularities in the orientation field are termed topological defects (see BOX.~2). Defects can form diverse patterns grouped into distinct classes based on the degree of rotation of directors around the defect cores (BOX.~2). In the absence of confinements and external topological constraints the defects with the lowest degree of distortions for nematic materials are the half-integer, $+1/2$ (comet-like) and $-1/2$ (three-fold symmetric), defects (see Fig.~\ref{fig:pol-nem}a) and are energetically favored over higher order defects as defects come with elastic energy~\cite{thijssen2020binding,vafa2020defect}. On the other hand, to minimize the elastic energy, polar materials (see Fig.~\ref{fig:pol-nem}a) favor formation higher order full-integer defects, i.e. $+1$ and $-1$ defects. Although, they mark imperfections in the alignment, topological defects are emerging as important organisation centres of cell cytoskeleton. A striking example is the full-integer polar topological defects in the form of aster-like structures in microtubule orientation field, which determine the location of bipolar spindles in the cell~\cite{roostalu2018determinants}. Moreover, in agreement with the predictions from models of active liquid crystals~\cite{metselaar19}, recent {\it in vivo} experiments have revealed topological defects in the actin filaments of the regenerating animal {\it hydra} as the organisation centres for animal morphogenesis, showing that full- and half-integer topological defects determine the location of foot, mouth, and tentacles of the animal during regeneration~\cite{maroudas2020topological}.\\

\noindent
\fcolorbox{red}{white}{\begin{minipage}{\textwidth}
{
{\bf BOX~2. Topological defects}\\
Topological defects mark singular points in the orientation field of the particles. They are topological because no smooth rearrangement of the particles can remove them, and they are defects because they correspond to imperfections in the alignment of particles, where domains with alignment mismatch meet each other. In two-dimensional orientation fields different kinds of topological defects can be formed that are distinguished based on their charge, which determines how much orientation of particles changes when a full rotation around the defect point is considered. For example, if during a full clock-wise turn ($+360$ degrees) around the defect point, orientation of particles rotates by $\pm 180$ degrees, that defect has an half-integer charge of $\pm 1/2$ ($\pm 180/+360$), while $\pm 360$ degrees rotation of particle orientations would correspond to a full-integer defect with $\pm 1$ charge (see Fig.~\ref{fig:pol-nem}a-c).\\
In unconfined, two-dimensional systems with nematic symmetry formation of full-integer defects is energetically costly and half-integer defects preferentially form. On the other hand two-dimensional polar systems allow full-integer defects. As such the charge of the defect can be considered as an indicator of the prevalent symmetry of that system~\cite{Doost18,Needleman2017}.\\
In active systems with nematic symmetry, the $+1/2$ defects are motile and their direction of motion can be used as an indicator of the form of forces created by active particles. For extensile particles that exert pushing forces along their elongation direction, $+1/2$ defects move towards their comet head, while $+1/2$ defects in contractile systems, which exert pulling forces towards their elongation direction, move towards their comet tail (Fig.~\ref{fig:pol-nem}d)~\cite{Needleman2017,Saw2018Review}.   
}
 \end{minipage}}
~\\

The structural form of topological defects is additionally important in determining the dominant symmetry of the system: while full-integer topological defects imply polar symmetry, the appearance of half-integer topological defects in the form of comet- and trefoil-like structures is the hallmark of the dominant nematic symmetry in any given system with orientational order (Fig.~\ref{fig:pol-nem}a). Indeed, both full- and half-integer topological defects have been reported in subcellular constructs comprising microtubule-kinesin motor mixtures~\cite{Guillamat17,roostalu2018determinants,opathalage2019self}, and actin filaments-myosin motor assemblies~\cite{kumar2018tunable}. Strikingly, the activity of these subcellular constructs renders topological defects as hotspots of flow generation. This is best characterised by the velocity field around half-integer topological defects in active nematics~\cite{Decamp2015,Zhang2021NatureMat}, that show stark contrast with passive liquid crystals. As such, because of their comet-like shape and continuous injection of energy by activity, positive half-integer defects can self-propel in active systems. This self-propulsion feature of comet-like defects recently revealed the role of defects in establishing chaotic dynamics at low Reynolds number due to the active nature of biological fluids~\cite{tan2019topological}. Moreover, the defect shape can also be altered by the activity of the system. A study on actin filaments and myosin motors~\cite{kumar2018tunable} revealed that the elastic properties of this active nematic system can be tuned by increasing the activity, leading to a decrease of the bending elasticity and changes in +1/2 defect morphology. 
 
\section*{Emergence of Liquid Crystalline Features in Cell Monolayers}
The power of applying concepts from physics to biological processes is in the universality of physical predictions. The physics of liquid crystal is no exception to this rule. As such, several liquid crystalline features found in subcellular filaments and cytoskeletal constructs within individual cells have been identified within cellular tissues that operate at completely different time and length scales compared to subcellular filaments. Here, in the tissue, the orientation field corresponds to the elongation direction of cells as a result of the anisotropic shape of the cells - either due to a naturally elongated shape such as in fibroblast~\cite{duclos2018spontaneous}, aggressive human breast cancer cell (MDA-MB-231)~\cite{Wullkopf18,PerezGonzales2019} or because of the cell deformation due to forces from neighbouring cells such as in MDCK and MCF-7 cells~\cite{Saw17,Balasubramaniam2021}.

\newpage

\subsection*{{\it Polar or Nematic Cells in Collective Cell Migration}}
There is now a growing list of cellular systems for which topological defects and liquid crystalline order have been identified. This includes epithelial cells such as Madine-Darby Canine Kidney (MDCK) cells, human skin keratinocyte (HaCaT) cells, luminal human breast cancer cells (MCF-7)~\cite{Saw17,Balasubramaniam2021}, Human Bronchial Cells (HBC)~\cite{Blanch18}, human breast cancer cells (MDA-MB-231)~\cite{PerezGonzales2019}, mouse myoblasts (C2C12)~\cite{guillamat2020integer}, human retinal epithelial cells (RPE1)~\cite{duclos2018spontaneous}, human fibrosarcoma (HT1080)~\cite{yashunsky2020chiral}, and neural progenitor stem cells~\cite{Kawaguchi17,yamauchi2020chirality}. Interestingly, with the exception of spatially confined monolayers in which a full-integer topological charge is imposed by the circular confinement~\cite{guillamat2020integer}, all the existing works have reported half-integer (nematic) topological defects in cell monolayers. Similarly, a three-dimensional reconstruction of the orientation field in mouse liver tissues has also identified nematic order in the cellular alignment~\cite{morales2019liquid}. These emergent nematic features - and the absence of polar features such as full-integer defects - in cell layers are particularly alluring since they suggest the generic predominance of nematic rather than polar symmetry in cellular monolayers. It is well-established that at an individual level cells are often endowed with polarity, yet, when multicellular layers are formed nematic features arise. Recent cell-based models of monolayers have suggested a possible explanation for this mismatch between polar symmetry at the single cell level and the nematic symmetry at multicellular level as the dominance of nematic alignment between the cells due to cell-cell adhesion: rather than aligning their polarity, upon cell-to-cell contacts neighboring cells simply deform such that they are elongated along a same axis~\cite{mueller19,Balasubramaniam2021}. Similar mechanistic description has been also proposed in theoretical models of cell layers based the physics of liquid crystals~\cite{amiri2021half}. Nevertheless, it is not completely clear if such a mismatch in symmetry or the emergent nematic features serve any physiological significance. Interestingly, collective cell migration can also result from polar order at the single cell level within multicellular clusters under certain conditions as exemplified by the formation of multicellular structures at the front of migrating tissues ~\cite{Vishwakarma:2018} or during collective cellular rotation {\it in vivo} or {\it in vitro}~\cite{Barlan2017,Jain2020}.  Overall, the interplay between polar and nematic orders in cellular assemblies remains unclear and could be a property reminiscent of the one observed at the scale of cytoskeleton inside a single cell within various actin structures. The ability to switch between polar and nematic organizations may contribute to the mechanical plasticity of tissues~\cite{JAIN2021}.

\subsection*{{\it Biological Functions of Active Nematics in Tissues}}
\begin{figure}[h]
\centering
  \includegraphics[trim={0 0 120 0},clip,width=1.0\linewidth]{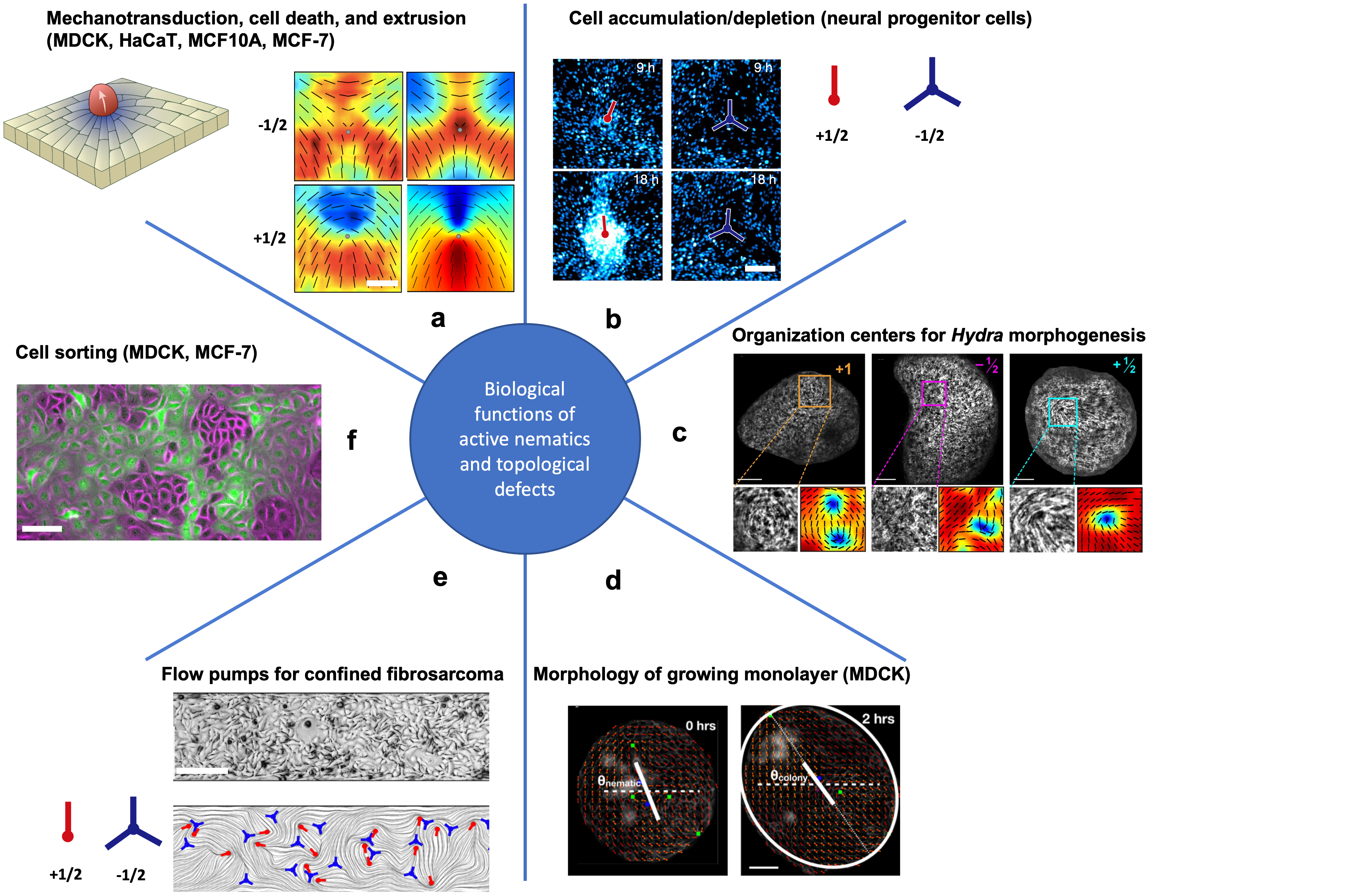}
  \caption{Biological significance of topological defects. Figures in (a)-(f) are adapted from~\cite{Saw17,Kawaguchi17,maroudas2020topological,comelles2021epithelial,yashunsky2020chiral,Balasubramaniam2021}, respectively. In (a) the colormaps represent isotropic stresses, ranging from negative compressive stresses (blue) to positive tensile stresses (red), overlaid on the director field (solid black lines) for experiments on MDCK cells (left column) and continuum model of the monolayer (right column). The scale bars in (a)-(f) are $50\mu m$, $100\mu m$, $100\mu m$, $50\mu m$, $200\mu m$, and $100\mu m$, respectively.}
  \label{fig:defects}
\end{figure}

Concerning the active nematic behavior of biological tissues, recent studies have definitely demonstrated the high potential of such approaches to better undersatnd biological processes. For instance, topological defects have been described as mechanotransduction hotspots (Fig.~\ref{fig:defects}a). Comet-like, positive half-integer defects govern cell death and consequent cell extrusion from epithelial monolayers (MDCK, MCF-7, HaCaT)~\cite{Saw17}. Modeling and experiments have demonstrated that compressive mechanical stresses build up at the head of the comet-shaped defect, resulting in the translocation of the transcriptional co-activator YAP to the cytoplasm. Since YAP suppresses pro-apoptotic genes, deactivation of YAP by compression may result in the activation of cell death signal (Caspase-3) and simultaneous extrusion of cells from the tissue. Remarkably, it is further shown that inducing comet-like defects by engineering the topology of the cell layer can be employed to localise cell extrusion events.
In addition, topological defects have been described as centers of cellular accumulation/depletion (Fig.~\ref{fig:defects}b). Experiments on neural progenitor stem cells have reported accumulation of cells at comet-like, positive half-integer defects and their depletion from trefoil-like, negative half-integer defects~\cite{Kawaguchi17}. It is further shown that cellular accumulation results in three-dimensional mound formation at the comet-like defects. The mechanism for this accumulation/depletion has been associated with anisotropic friction in cell layers, easing cellular motion along the nematic axis and resisting perpendicular motion to it. Recently it has been demonstrated that the change in density of cells around topological defects in mesothelium can result in the suppression of the ovarian cancer cells clearance from the mesothelium~\cite{zhang2021topological}. Furthermore, recent experiments in confined myoblasts have also demonstrated the formation of three-dimensional helical mounds at spiral full-integer defects that are formed by merging two half-integer defects due to the imposed circular confinement~\cite{guillamat2020integer}.

Importantly, a large number of non-equilibrium biological processes take place in confined environments. This is particularly true during tissue morphogenesis~\cite{MORITA2017}. As such, boundary effects may influence the behaviours of active nematics. In particular, a spherical topology ensures that the total charge
of the surface-bound topological defects is $+2$. Along this line, a recent study has shown that during hydra regeneration, the system folds into spheroids with a total charge of $+2$. Importantly, the emergence of two $+1$ topological defects during Hydra morphogenesis visualized by the pluricellular actin organization coincides with the apex of the head and the base of the foot, thus defining the body axis of the animal~\cite{maroudas2020topological}. The further development of the animal leads to the formation of tentacles that are again associated with topological defects of charge $+1$ at the tip leaving behind two $-1/2$ at the base (Fig.~\ref{fig:defects}c), that is consistent with earlier predictions from continuum modeling of active nematic shells~\cite{metselaar19}.

In line with earlier predictions from models of active liquid crystals~\cite{Doostmohammadi16}, recent experiments on epithelial cells (MDCK) have shown that the self-propulsion of comet-like defects are correlated with the protrusion formation and elongation of initially circular cell colonies (Fig.~\ref{fig:defects}d)~\cite{comelles2021epithelial}. Under cellular confinement, topological defects have been identified in cellular monolayers as pumps of cellular flows. Indeed, recent experiments on confined human fibrosarcoma have reported, in agreement with models of active liquid crystals, the emergence of chiral currents of cells at the edges of confinement, identifying comet-like defects close to the boundaries as the sources of cell currents (Fig.~\ref{fig:defects}e)~\cite{yamauchi2020chirality}.

Finally, topological defects and nematic activity can reflect the collective cellular self-organisation modes. It is recently found that weakening cell-cell adhesion by removing E-cadherin protein (E-cad KO) in epithelial cells (MDCK and MCF-7) results in the switching of the self-propulsion direction of comet-like defects, upending their behavior from extensile (see Fig.~\ref{fig:pol-nem}d and BOX 2) in wild-type layers to contractile in E-cad KO cells (Fig.~\ref{fig:defects}f)~\cite{Balasubramaniam2021}. The mechanism for this switch was explained based on the competition between contractile cell-substrate stresses and extensile cell-cell contact stresses, showing that sufficient reduction of cell-substrate interaction in E-cad KO cells by adding blebbistatin or culturing cells on soft substrates rescues the extensile behavior. Since the loss of E-cadherin is a hallmark of epithelial-to-mesenchymal transition (EMT) the dynamics of mixtures of wild-type and E-cad KO cells was explored showing autonomous cell sorting governed by the competition between extensile and contractile stresses~\cite{Balasubramaniam2021} and providing alternative mechanisms to the existing models of differential adhesion or tension~\cite{FOTY2005,Maitre2012,Sahu2020Sorting}.

\section*{Exploiting Physics of Liquid Crystals for Guiding Cellular Self-organisation}
The increasingly emerging physics of liquid crystals in cellular systems and the growing list of biological functionalities for topological defects, provide a fertile ground for exploiting physics of liquid crystals in designing advanced biomaterials for improved tissue regeneration and for biomimetic inspirations. Beyond the applications of liquid crystals in optical biosensors and actuators~\cite{woltman2007liquid}, the design and engineering of liquid crystal based biomaterials for tissue repair and regeneration is still in its infancy. Nevertheless, recent experiments have provided early evidence on applicability of the physics of liquid crystal principles in controlling cellular organisation. It is shown that surfaces pre-patterned with liquid crystal elastomers (LCE) can be used for controlling the orientation of human dermal fibroblasts (HDF)~\cite{turiv2020topology}. Upon swelling in an aqueous medium, the elastomers form nano-scales ridges that guide alignment of elongated fibroblast cells in a monolayer. Using this approach it is shown that pre-designed patterns of half- and full-integer defects can be employed to induce accumulation and depletion centers for cells, suggesting a means of controlling cell density in space. Moreover, such a pre-imposed pattern is shown to be useful in stabilising cellular orientation field to allow inferring elastic properties of the cell monolayers based on theories of liquid crystals. Similar inference of cell layer material properties based on theories of liquid crystals are also proposed by confining myoblasts in small circular adhesive islands to stabilise spiral topological defects at the centre of the confinement~\cite{blanch2021quantifying}. More recently, it is further shown that pre-patterned micron-sized ridges can be exploited to control alignment of fibroblast (3T6) and epithelial (EpH-4) cell layers to from stable full-integer topological defects~\cite{endresen2021topological}. Remarkably, it was shown that such pre-imposed patterns can control the activation/deactivation of mechanotransduction by triggering nuclear/cytoplasmic YAP translocation at topological defects with positive and negative charges and in different cell types. 
 
\section*{Concluding Remarks}
In this review, we provide examples of active nematic liquid crystals, highlighting how they drive self-organization of biological systems at multiple length-scales. Nematic ordering and topological defects have been identified in large variety of systems including cytoskeletons, various cell lines and tissues and even others such as bacterial colonies not reviewed here ~\cite{Meacock2021,Copenhagen2021}. From these studies, it emerges that topological defects are hotspots of mechanical signals ~\cite{Saw17,Kawaguchi17}. They can also serve as organization centers to shape tissues and organisms as shown in the Hydra development~\cite{maroudas2020topological}. In addition, recent studies have highlighted the capacity of using LCE to control the organization of multicellular assemblies. Such approaches may help to further understand the interplay between extra-cellular matrix properties and tissue dynamics. Further understanding of liquid crystal-based approaches  will also require a more profound insight into polar versus nematic organisation at subcellular and multicellular scales. How can both structures coexist? Could biological systems switch from one structure to the other? How such dynamics may impact on single cell polarity through cytoskeleton regulation as well as on large-scale polarization of tissues? Additionally, cell layers establish collective patterns of motion on length scales that are often larger than the scale of individual cells~\cite{lachance2021learning}. It is not clear how these collective migration length scales are correlated with length scales associated with orientational order and liquid crystalline features in the tissue. We anticipate that answering these questions will shed new light on developmental and pathological processes.
Additionally, majority of studies so far have focused on liquid-crystalline features in two-dimensional systems. Recent experiments on reconstituted microtubule-motor protein mixtures stabilised by filamentous viruses have provided first {\it in vitro} realisation of three-dimensional active topological defects in the form of lines and loop~\cite{duclos2020topological}, as predicted from the theory of active nematics~\cite{shendruk2018twist,vcopar2019topology}. The existence and possible functionality of such three-dimensional features in cell assemblies presents an intriguing venue for further studies of active liquid crystals in cell biology.
Finally, an interesting perspective to further understand tissue mechanics concerns the links between active nematics and the physics of jamming-unjamming~\cite{SADATI2013}. Similar parameters such as cell density, shape, cell-cell and cell-matrix adhesions ~\cite{Garcia2015,Balasubramaniam2021,Bi2015} are at play in both descriptions but the interplay between these approaches remains to be established. Particularly, it is suggested that the jamming-unjamming transition coincides with a topological transition in cell shape as the ratio of the cell perimeter to the square-root of its area exceeds a threshold~\cite{Atia2018}. Interestingly, recent theoretical work ~\cite{mueller19} suggests that the emergence of orientational order and topological defects is also correlated with the cellular shape changes leading to the elongation of otherwise isotropic circular cells. It is an attractive idea to connect such topological transitions in cell shape changes to the cellular elongation and the establishment of orientational order in cell aggregates.\\

\noindent
\fcolorbox{red}{white}{\begin{minipage}{\textwidth}
{
{\bf Highlights}
 \begin{itemize}
     \item Various forms of liquid crystalline order including nematic, smectic and chiral features have been established in cytoskeletal constructs {\it in vitro} and in subcellular filaments {\it in vivo}.
     \item Nematic ordering and topological defects are identified in an unprecedented speed in tissues of various cell types.
     \item Topological defects are emerging as hotspots of mechanotransduction and organisation centres for cellular flows in tissues.
     \item Topological design through pre-patterned surfaces and geometric constraints is appearing as a promising tool for exploiting physics of liquid crystals for guiding cellular self-organisation.
 \end{itemize}   
}
 \end{minipage}}
~\\

\noindent
\fcolorbox{red}{white}{\begin{minipage}{\textwidth}
{
{\bf Outstanding Questions}
  \begin{itemize}
     \item Can passive liquid crystals be used to pre-organize biological tissues?
     \item Can polar and nematic organizations coexist in collective cell behaviors?
     \item What are the links between the physics of liquid crystals and the physics of jamming-unjamming?
     \item What could be the role and structure of topological defects in 3D cellular tissues and multilayered assemblies?
     \item How can topological defects influence {\it in vivo} morphogenetic processes?
 \end{itemize}   
}
 \end{minipage}}
\section*{Acknowledgement and Funding}
We acknowledge support from the Novo Nordisk Foundation (grant no. NNF18SA0035142), Villum Fonden (grant no. 29476), funding from the European Union’s Horizon 2020 research and innovation program under the Marie Sklodowska-Curie grant agreement no. 847523 (INTERACTIONS), the LABEX  Who Am I? (ANR-11-LABX-0071), the Ligue Contre le Cancer (Equipe labellisée 2019), and the Agence Nationale de la Recherche (‘POLCAM’ ANR-17-CE13-0013 and ‘MyoFuse’ ANR-19-CE13-0016). We thank Julia Yeomans, Thuan Beng Saw and the group members from ``Cell adhesion and migration" team and from ``Active \& Intelligent Matter'' group for helpful discussions. 
\bibliographystyle{apsrev4-1}
\bibliography{references}
\end{document}